\documentclass[conference]{IEEEtran}

\ifCLASSINFOpdf
\else
\fi

\usepackage{cleveref}
\usepackage{amsmath,amssymb}
\usepackage{enumerate}
\usepackage{amsfonts}
\usepackage{mathtools}
\usepackage{tikz}
\usepackage{cite}
\usepackage{lipsum}
\usepackage{amsfonts}
\usepackage{graphicx}
\usepackage{epstopdf}
\usepackage{algorithmic}
\usepackage{amsopn}

\newtheorem{example}{Example}

\newtheorem{proposition}{Proposition}

\newtheorem{definition}{Definition}
\newtheorem{theorem}{Theorem}
\newtheorem{lemma}{Lemma}

\newcommand{\F}{\mathbb{F}}
\newcommand{\K}{\mathbb{K}}

\DeclareMathOperator{\rep}{Rep}

\tikzstyle{zcorner}  = [draw, ellipse, minimum size=10pt, align=center]

\begin{document}

	\title{Private Computation of Systematically Encoded Data with Colluding Servers}
	
\author{\IEEEauthorblockN{David Karpuk}
\IEEEauthorblockA{Departamento de Matem\'aticas\\
Universidad de los Andes\\
Bogot\'a, Colombia\\
Email: da.karpuk@uniandes.edu.co}}

	\maketitle

	\begin{abstract}
	Private Computation (PC), recently introduced by Sun and Jafar, is a generalization of Private Information Retrieval (PIR) in which a user wishes to privately compute an arbitrary function of data stored across several servers.  We construct a PC scheme which accounts for server collusion, coded data, and non-linear functions.  For data replicated over several possibly colluding servers, our scheme computes arbitrary functions of the data with rate equal to the asymptotic capacity of PIR for this setup.  For systematically encoded data stored over colluding servers, we privately compute arbitrary functions of the columns of the data matrix and calculate the rate explicitly for polynomial functions.  The scheme is a generalization of previously studied star-product PIR schemes.
	\end{abstract}

	\IEEEpeerreviewmaketitle

	\section{Introduction}

	\subsection{Background and Summary of Current Results}
	
	The problem of Private Information Retrieval (PIR), introduced in \cite{PIR_original}, is to download a file from a database without revealing the identity of the file.  The data is usually assumed to be encoded over several servers, subsets of which may collude to try to deduce the identity of the file.  The \emph{rate} of a PIR scheme is the size of the desired file to the total amount of data the user must download, and the \emph{capacity} of a given PIR setup is the maximum possible rate.
	
	The capacity of PIR for data replicated over several servers was obtained in \cite{sun_jafar_1}, for replicated data with colluding servers in \cite{sun_jafar_2}, and for MDS-coded data without server collusion in \cite{bananaman}.  The capacity of PIR for MDS-coded data with server collusion remains open, but the star-product scheme of \cite{FGHK16} has, as far as the author knows, the best rates yet obtained as the number of files goes to infinity. Related PIR schemes and capacity results have been studied in\cite{patternISIT2017,zhang_ge_variants,kumar_PIR_journal,us_rm_pir,bananamanbyz}.
	
	Private Computation (PC), introduced by Sun and Jafar in \cite{sunjafarpc}, generalizes PIR by allowing the user to compute arbitrary linear combinations of the files for replicated data.  Surprisingly, the capacity of this setup is equal to the PIR capacity of \cite{sun_jafar_1}.  This leaves open a multitude of questions, most notably scheme constructions for PC which account for colluding servers, coded data, non-linear functions, etc.
	
	The current contribution generalizes the star-product scheme of \cite{FGHK16} to construct explicit PC schemes for colluding servers, coded data, and non-linear functions.  In particular, we show that for a data vector replicated over $N$ servers with $T$-collusion, private computation of arbitrary functions is possible with download rate $(N-T)/N$, which is the asymptotic capacity of PIR with $T$-collusion.  For data encoded using an $[N,K]$ systematic storage code, we construct a scheme which evaluates arbitrary functions of the columns of the data matrix and protects against $T$-collusion.  When using Reed-Solomon codes and computing polynomial functions of degree $G$, our scheme has rate $\min\{N-(G(K-1)+T),K\}/N$.  While the rates we obtain are highly suggestive of asymptotic capacity results, we sidestep capacity questions in favor of explicit constructions.

	\subsection{Some Coding-Theoretic Preliminaries}
	
	We assume basic familiarity with linear coding theory, and refer to \cite{ma77} as a general reference.  If $\mathcal{C}$ is a linear code over a finite field $\F$ with length $N$, dimension $K$, and minimum distance $D$, we refer to it as an $[N,K,D]_\F$ code, and possibly omit the $D$ or $\F$ if they are irrelevant or clear from context.  A generator matrix for $\mathcal{C}$ is denoted by ${\bf G}_\mathcal{C}\in \F^{K\times N}$, and a parity-check matrix by ${\bf H}_{\mathcal{C}}\in \F^{N\times (N-K)}$.  Recall that $\mathcal{C}$ is MDS if and only if $D = N-K+1$ if and only if every $K$ columns of ${\bf G}_\mathcal{C}$ are linearly independent.  The repetition code over $\F$ is denoted by $\rep(N)_\F$, and it is an $[N,1]$ MDS code.
	
	
	
	\subsubsection{Star Products}
	
	Let $\mathcal{C}$ and $\mathcal{D}$ be linear codes of length $N$.  Their \emph{star product} $\mathcal{C}\star \mathcal{D}\subseteq\F^{1\times N}$ is the $\F$-span of all vectors of the form
	\begin{equation}
	{\bf c}\star {\bf d} = \left[{\bf c}(1){\bf d}(1),\ldots,{\bf c}(N){\bf d}(N)\right]\in \F^{1\times N}.
	\end{equation}
	for ${\bf c}\in \mathcal{C}$ and ${\bf d}\in \mathcal{D}$.  For any $G>0$, the \emph{$G$-fold star product} of $\mathcal{C}$ with itself is
	\begin{equation}
	\mathcal{C}^{\star G} = \text{span}_\F\{{\bf c}_1\star \cdots \star {\bf c}_G\ |\ {\bf c}_g\in \mathcal{C}\}
	\end{equation}
	Note that $\mathcal{C}\star\rep(N)_\F = \mathcal{C}$ for any linear code $\mathcal{C}$, and that if $\rep(N)_\F\subseteq \mathcal{C}$ then $\mathcal{C}\subseteq\mathcal{C}\star\mathcal{C}$ and thus by induction $\mathcal{C}^{\star G_1}\subseteq\mathcal{C}^{\star G_2}$ for any $G_1\leq G_2$.
	
	\subsubsection{Reed-Solomon Codes}
	
	Let $\boldsymbol{\alpha} = \left[\boldsymbol{\alpha}(1),\ldots,\boldsymbol{\alpha}(N)\right]\in \F^{1\times N}$ be such that $\boldsymbol{\alpha}(i)\neq \boldsymbol{\alpha}(j)$ for all $i\neq j$.  For any $K\leq N$, we define the $[N,K]$ \emph{Reed-Solomon Code} (RS code) associated to this data by
	\begin{equation}
	\mathcal{RS}_K(\boldsymbol{\alpha}) = \{
	\boldsymbol{\phi}(\boldsymbol{\alpha})\ |\ 
	\phi\in \F[X],\ \deg(\phi)<K\}
	\end{equation}
	where $\boldsymbol{\phi}(\boldsymbol{\alpha}) = \left[\phi(\boldsymbol{\alpha}(1)),\ldots,\phi(\boldsymbol{\alpha}(N))\right] \in \F^{1\times N}$.  We call $\boldsymbol{\alpha}$ the \emph{evaluation vector} of $\mathcal{RS}_K(\boldsymbol{\alpha})$.  It is well-known that $\mathcal{RS}_K(\boldsymbol{\alpha})$ is an $[N,K]$ MDS code.  For two RS codes with the same evaluation vector, we have the following:
	\begin{equation}\label{rsstar1}
	\mathcal{RS}_K(\boldsymbol{\alpha}) \star \mathcal{RS}_L(\boldsymbol{\alpha}) = \mathcal{RS}_{\min\{K+L-1,N\}}(\boldsymbol{\alpha})
	\end{equation}
	for which we refer to \cite{FGHK16} for a proof.  By induction we have
	\begin{equation}\label{rsstar2}
	\mathcal{RS}_K(\boldsymbol{\alpha})^{\star G} = \mathcal{RS}_{\min\{G(K-1)+1,N\}}(\boldsymbol{\alpha}).
	\end{equation}

	\subsubsection{Base Field Extension}
	
	Let $\mathcal{C}\subseteq\F^{1\times N}$ be a linear code, and let $\K/\F$ be a finite field extension of $\F$.  We define the \emph{extension of $\mathcal{C}$ to $\K$} to be the code $\mathcal{C}_\K\subseteq \K^{1\times N}$ given by the $\K$-span of the rows of ${\bf G}_\mathcal{C}$.  Base field extension does not change any essential properties of the code:
	\begin{proposition}
	Let $\mathcal{C}$ be an $[N,K,D]_\F$ code, and let $\K/\F$ be a finite field extension.  We have the following:
	\begin{itemize}
	\item[(i)] $\mathcal{C}_\K$ is an $[N,K,D]_\K$ code, hence if $\mathcal{C}$ is MDS so is $\mathcal{C}_\K$. 
	\item[(ii)] The matrices ${\bf G}_\mathcal{C}$ and ${\bf H}_\mathcal{C}$ are generator and parity-check matrices, respectively, for $\mathcal{C}_\K$.
	\item[(iii)] For any code $\mathcal{D}\subseteq\F^{1\times N}$, we have $(\mathcal{C}\star \mathcal{D})_\K = \mathcal{C}_\K\star \mathcal{D}_\K$.
	\item[(iv)] $\mathcal{RS}_K(\boldsymbol{\alpha})_\K$ is also an RS code over $\K$, where we view the vector ${\boldsymbol{\alpha}}$ as an element of $\K^{1\times N}$.
	\end{itemize}
	\end{proposition}
	\begin{IEEEproof}
	The proofs of all of these statements follow immediately from the definitions.
	\end{IEEEproof}

	\section{System Model and Problem Statement}

	\subsection{Basic Definitions}\label{defns}
	
	Let $\F$ be a finite field, and let $\K/\F$ be a finite field extension of $\F$.  We consider a data matrix
	\begin{equation}
	{\bf X} = \left[
	{\bf x}_1 \cdots  {\bf x}_K
	\right]
	\in \K^{M\times K}
	\end{equation}
	consisting of $K$ column vectors ${\bf x}_k\in \K^{M\times 1}$.  The matrix ${\bf X}$ is stored across $N$ servers using an $[N,K]_\F$ storage code $\mathcal{C}$ by defining
	\begin{equation}
	{\bf X}\cdot {\bf G}_\mathcal{C} = {\bf Y} = \left[
	{\bf y}_1 \cdots {\bf y}_N
	\right]\in \K^{M\times N}
	\end{equation}
	Server $n$ now stores the $n^{th}$ column vector ${\bf y}_n\in \K^{M\times 1}$.

	It will be convenient to define
	\begin{equation}
	\mathcal{F} = \{\text{all functions } \K^{M\times 1}\rightarrow \K\}
	\end{equation}
	as the space of all possible queries.  Given some $B$ functions $\phi_b\in \mathcal{F}$, the goal is to privately compute all values $\phi_b({\bf x}_k)$, for $1\leq b\leq B$ and $1\leq k\leq K$.  We refer to $B$ as the \emph{block length}, and view it as a flexible parameter the user can adjust to download the $\phi_b({\bf x}_k)$ with maximal efficiency.

	\begin{definition}
	Given a data matrix ${\bf X}\in \K^{M\times K}$, an encoded data matrix ${\bf X}\cdot {\bf G}_\mathcal{C} = {\bf Y}\in \K^{M\times N}$, and a block length $B$, a \emph{Private Computation} (PC) scheme for this setup consists of:
	\begin{enumerate}
	\item A \emph{query space} $\mathcal{Q}\subseteq \mathcal{F}$, from which $B$ functions $\phi_b$ to be evaluated on the columns of ${\bf X}$ are sampled according to a random variable $\mathsf{Q}$ on $\mathcal{Q}^B$.  Both $\mathcal{Q}$ and $\mathsf{Q}$ are made public.
	\item \emph{Queries} $\rho_1,\ldots,\rho_n\in \mathcal{Q}$, where $\rho_n$ is transmitted by the user to the $n^{th}$ server.
	\item \emph{Responses} $\rho_n({\bf y}_n)\in \K$, computed by the servers and transmitted back to the user, who receives the \emph{total response vector}
	\begin{equation}
	\boldsymbol{\rho}({\bf Y}) = \left[ \rho_1({\bf y}_1), \ldots,  \rho_N({\bf y}_N) \right] \in \K^{1\times N}
	\end{equation}
	\item An \emph{iteration process}, wherein the user repeats steps 2.\ and 3.\ a total of $S$ times (choosing different queries during each iteration) until all function values $\phi_b({\bf x}_k)$ can be computed from all of the $S$ vectors $\boldsymbol{\rho}({\bf Y})$.
	\end{enumerate}
	\end{definition}
	
	Our main measurement of the efficiency of a PC scheme is the download rate, defined as follows.
	
	\begin{definition}
	Given a PC scheme with parameters as in the previous definition, the \emph{PC rate}, or \emph{download rate}, or simply \emph{rate} of the PC scheme is defined to be
	\begin{equation}
	R = \frac{KB}{NS}.
	\end{equation}
	\end{definition}
	
	The assumption that $[\K:\F]\gg0$ generally allows one to ignore upload costs and focus on the download rate of a scheme as the primary performance metric.  Nothing is lost mathematically in the construction and analysis of our schemes, however, by assuming that $\K = \F$.
	
	\begin{definition}
	Let $\mathcal{T}\subseteq [N]$ be a subset of servers of size $T$.  A PC scheme \emph{protects against the colluding set $\mathcal{T}$} if
	\begin{equation}\label{mutinfo}
	I(\mathsf{Q};\mathsf{P}_\mathcal{T}) = 0
	\end{equation}
	where $\mathsf{P}_\mathcal{T}$ is the random variable given by all queries sent to all servers in $\mathcal{T}$ over all $S$ iterations of the scheme.  If a PC scheme protects against every subset of servers of size $T$, we say that the scheme \emph{protects against $T$-collusion}, or is \emph{$T$-private}, and refer to it as a \emph{$T$-PC scheme}.
	\end{definition}

	\subsection{Comparison with Other Private Computation Models and Private Information Retrieval}\label{compwithPIR}
	
	We now briefly discuss how our model of Section \ref{defns} compares with two previous works on Private Computation, notably \cite{sunjafarpc,miropc}.  Assume $\K = \F$ for simplicity.  In \cite{sunjafarpc,miropc} the authors set $K = 1$ and consider a replication system with $\mathcal{C} = \rep(N)_\F$, so each of $N$ servers is storing a copy of ${\bf X}\in \F^{M\times 1}$.  The vector ${\bf X}$ is divided into $M'$ datasets ${\bf X}^{m'}\in \F^{B\times 1}$, so that $M = BM'$, and the user wishes to compute a linear combination
	\begin{equation}
	\lambda_1{\bf X}^1 + \cdots +\lambda_{M'}{\bf X}^{M'}\in \F^{B\times 1},\quad \lambda_{m'}\in \F
	\end{equation}
	of the datasets.  This is equivalent to computing the matrix multiplication $\boldsymbol{\Phi}\cdot {\bf X}$, where $\boldsymbol{\Phi}\in \F^{B\times M}$ is the block matrix
	\begin{equation}
	\boldsymbol{\Phi} = \left[
	\lambda_1{\bf I}_B \cdots \lambda_{M'}{\bf I}_B
	\right]
	\end{equation}
	Setting the $\phi_b$ to be the linear functions defined by the rows of $\boldsymbol{\Phi}$ then reproduces the main problem of \cite{sunjafarpc,miropc} as an particular instance of our model (here one can set $\mathcal{Q}$ to be the set of all linear functions $\F^{M\times 1}\rightarrow \F$, and $\mathsf{Q}$ to be uniform on some subset of all such above $\boldsymbol{\Phi}$).  In addition to generalizing the model of \cite{sunjafarpc} from certain linear functions of the data vector to arbitrary functions, our model is perhaps simpler in the sense that we avoid discussing datasets and discard the parameter $M'$.
	
	Private Information Retrieval (PIR) is simply a special instance of the above PC problem, in which some $\lambda_w = 1$ and $\lambda_{m'} = 0$ for $m'\neq w$.  In general, one considers arbitrary storage codes ($K>1$) as well as non-trivial server collusion ($T>1$), but substituting an arbitrary storage code $\mathcal{C}$ into the above discussion or accounting for server collusion does not affect the way in which PC generalizes PIR.

	\subsection{Private Polynomial Computation}
	
	Our motivating example will be computing polynomial functions of the data matrix.  More precisely, given a positive integer $G$, let
	\begin{equation}
	\mathcal{P}_G = \{\phi\in \F[X_1,\ldots,X_M]\ |\ \deg(\phi)\leq G,\ \phi({\bf 0}) = 0\}
	\end{equation}
	be the $\F$-vector space of polynomials in $M$ variables with coefficients in $\F$ and no constant term, whose total degree is bounded above by $G$.  Here we recall that the degree of a multivariate monomial $\phi = X_1^{a_1}\cdots X_M^{a_M}$ is defined to be $\deg(\phi) = \sum_m a_m$, and for a general polynomial, $\deg(\phi)$ is the maximal degree of all the monomials appearing in $\phi$.  
	

	\begin{proposition}\label{polyfuns}
	Consider the $\F$-linear map $\eta:\mathcal{P}_G\rightarrow \mathcal{F}$ taking a polynomial to the function it defines.  If $G<|\K|$, then this map is injective.
	\end{proposition}
	\begin{IEEEproof}
	We have to show that $\phi$ does not define the zero function on $\K^{M\times 1}$, for all $\phi\in \mathcal{P}_G$.  Let ${\bf x}\in \K^{M\times 1}$ be selected uniformly at random.  By the Schwartz-Zippel Lemma, we have $\text{Pr}(\phi({\bf x})=0) \leq G/|\K| < 1$.  In particular, there exists an ${\bf x}$ such that $\phi({\bf x})\neq 0$, which proves the statement.
	\end{IEEEproof}
	
	
	To privately compute $\phi_b({\bf x}_k)$ where the functions $\phi_b$ are polynomials of degree bounded by $G$, we set the query space to be $\mathcal{Q} = \eta({\mathcal{P}}_G)$.  When $G< |\K|$, Proposition \ref{polyfuns} allows us to ignore the distinction between polynomials and the functions they define, and simply set $\mathcal{Q} = \mathcal{P}_G$.

	\section{$T$-Private Computation of Replicated Data}
	
	In this section we assume that $C = \rep(N)_\F$, so $K = 1$ and every server stores a copy of ${\bf X}\in \K^{M\times 1}$.  Thus the goal is to $T$-privately compute $\phi_b({\bf X})$ for some $B$ functions $\phi_b:\K^{M\times 1}\rightarrow \K$.  We show that for any query space $\mathcal{Q}$ which is a vector space over $\F$, we can perform this computation with download rate equal to $(N-T)/N$, which is the asymptotic capacity as $M\rightarrow\infty$ of $T$-PIR.
	
	\subsection{Scheme Construction}\label{scheme}

	We assume that $\mathcal{Q}\subseteq\mathcal{F}$ is a vector space over $\F$.  We do \emph{not} assume that the elements of $\mathcal{Q}$ are themselves linear functions, only that $\mathcal{Q}$ is closed under function addition and scalar multiplication by elements of $\F$.
	
	We set $Q = \dim_\F \mathcal{Q}$ and let $\{\psi^1,\ldots,\psi^Q\} \subset \mathcal{Q}$ be an $\F$-basis of $\mathcal{Q}$.  Our scheme is based on the construction of \cite{FGHK16} for replicated data.  In particular, we use an $[N,T]$ MDS \emph{retrieval code} $\mathcal{D}$ to guarantee that the queries sent to any $T$ servers are uniformly distributed on $\mathcal{Q}^T$ and independent of the functions $\phi_b$.  We retrieve the function values by decoding in $\mathcal{D}$.
	
	Let $\mathcal{D}$ be an $[N,T]_\F$ MDS code.  The first iteration of our scheme proceeds as follows.  We sample, independently and uniformly, codewords ${\bf d}^1,\ldots,{\bf d}^Q\in \mathcal{D}$.  For each server $n = 1,\ldots,N$, define
	\begin{equation}\label{psin}
	\psi_n = {\bf d}^1(n)\cdot \psi^1 + \cdots + {\bf d}^Q(n)\cdot \psi^Q \in \mathcal{Q}
	\end{equation}
	where ${\bf d}^q(n)\in \F$ denotes the $n^{th}$ coordinate of ${\bf d}^q$.  Suppose we want to evaluate functions $\phi_1,\ldots,\phi_B\in \mathcal{Q}$, sampled according to $\mathsf{Q}$.  We define the queries as follows:
	\begin{equation}
	\rho_n = \left\{\begin{array}{cl}
	\psi_n + \phi_n & \text{if $n\leq N-T$} \\
	\psi_n & \text{if $n> N-T$}
	\end{array}
	\right.
	\end{equation}
	The total response from the servers is given by
	\begin{equation}\label{totalresponse}
	\begin{aligned}
	\boldsymbol{\rho}({\bf X}) &= \underbrace{\left[\psi_1({\bf X}),\ldots,\psi_N({\bf X})\right]}_{=:\boldsymbol{\psi}({\bf X})} + \left[\phi_1({\bf X}),\ldots,\phi_{N-T}({\bf X}),{\bf 0}\right]
	\end{aligned}
	\end{equation}
	We claim that $\boldsymbol{\psi}({\bf X})\in \mathcal{D}_\K$.  To see this, we write
	\begin{align}
	\boldsymbol{\psi}({\bf X}) &= \sum_{q = 1}^Q \left[
	{\bf d}^q(1)\psi^q({\bf X}),
	\ldots,
	{\bf d}^q(N)\psi^q({\bf X})
	\right] \\
	&=
	\sum_{q = 1}^Q\psi^q({\bf X})\cdot {\bf d}^q \in \mathcal{D}_\K
	\end{align}
	Let ${\bf H}_\mathcal{D}\in \F^{N\times (N-T)}$ be a parity-check matrix of $\mathcal{D}$ in systematic form.  We retrieve the desired function evaluations by right-multiplying $\boldsymbol{\rho}({\bf X})$ by ${\bf H}_\mathcal{D}$.  Since $\boldsymbol{\psi}({\bf X})\in \mathcal{D}_\K$ we have that $\boldsymbol{\psi}({\bf X})\cdot {\bf H}_\mathcal{D} = {\bf 0}$, and hence
	\begin{equation}
	\boldsymbol{\rho}({\bf X})\cdot {\bf H}_\mathcal{D} = \left[
	\phi_1({\bf X}),
	\ldots,
	\phi_{N-T}({\bf X})
	\right]
	\end{equation}
	hence we retrieve the desired function evaluations.  
	
	Now suppose that $N-T$ divides $B$.  To compute the remaining function evaluations $\phi_{N-T+1}({\bf X}),\ldots,\phi_B({\bf X})$, we simply iterate the above procedure a total of $B/(N-T)$ times, each time choosing the next $N-T$ of the functions $\phi_b$ in the obvious way.  The scheme construction is complete.
	
	\subsection{Proofs of Correctness and $T$-Privacy}
	
	The $T$-privacy of the above scheme construction will follow from the following useful Lemma.
	
	\begin{lemma}\label{uniform}
	Let $\mathcal{V}$ be a finite-dimensional vector space over a finite field, and let $\mathsf{U}$ be a uniform random variable on $\mathcal{V}$.  Let $\mathsf{Z}$ be any other random variable on $\mathcal{V}$ which is independent of $\mathsf{U}$.  Then $\mathsf{W} = \mathsf{U} + \mathsf{Z}$ is uniform on $\mathcal{V}$ and $I(\mathsf{W};\mathsf{Z}) = 0$.
	\end{lemma}
	\begin{IEEEproof}
	This is a straightforward calculation using the pmfs of all three random variables. 
	\end{IEEEproof}

	\begin{theorem}\label{repthm}
	Suppose that $N-T$ divides $B$.  Then for any $\F$-linear query space $\mathcal{Q}$, the above PC scheme protects against $T$-collusion and has download rate $(N-T)/N$.
	\end{theorem}
	\begin{IEEEproof}
	The proof of $T$-privacy is analogous to that of the main scheme of \cite{FGHK16}.  Specifically, let $\mathcal{T}\subseteq[N]$ be a subset of servers of size $T$, and consider the projection $\mathcal{D}_\mathcal{T}$ of $\mathcal{D}$ onto the coordinates in $\mathcal{T}$.  Since $\mathcal{D}$ is MDS, this projection $\mathcal{D}\rightarrow \mathcal{D}_\mathcal{T} = \F^{1\times T}$ is surjective.  It follows that if $\mathcal{T} = \{n_1,\ldots,n_T\}$, then the tuple
	\begin{equation}
	\left[\psi_{n_1},\ldots,\psi_{n_T}\right]\in \mathcal{Q}^T
	\end{equation}
	defines the uniform random variable on $\mathcal{Q}^T$.  Since the $\phi_b$ and $\psi_n$ are chosen independently, Lemma \ref{uniform} shows that we protect against $T$-collusion during a single iteration.  As the source of randomness is independent between iterations, the condition (\ref{mutinfo}) follows in general.  Computing the download rate gives $R = \frac{B}{\frac{B}{N-T}\cdot N} = \frac{N-T}{N}$ as claimed.
	\end{IEEEproof}
	
	When $T = 1$ and $N>1$, the PC scheme of \cite{sunjafarpc} achieves a rate of $(1-1/N)/(1-(1/N)^M)$, which they show is the capacity for the query space $\mathcal{Q}$ of linear functions described in Section \ref{compwithPIR}.  The scheme of Section \ref{scheme} achieves the limiting value $(N-1)/N$ as $M\rightarrow \infty$, for any $\F$-linear query space whatsoever.  The capacity of $T$-PC is unknown for $T>1$ and arbitrary $\mathcal{Q}$, but it our hope that the rate $(N-T)/N$ we achieve is equal to asymptotic capacity, for any $\F$-linear $\mathcal{Q}$.
	

	\begin{example}\label{poleax}
	Suppose that $N = 3$ and $T = 2$.  To protect against $2$-collusion we set $\mathcal{D}$ to be the $[3,2,2]_{\F_2}$ code with
	\begin{equation}
	{\bf G}_\mathcal{D} = \begin{bmatrix}
	1 & 0 & 1 \\
	0 & 1 & 1
	\end{bmatrix}
	\end{equation}
	We take the block length to be $B = 1$ and set the query space to be $\mathcal{Q} = \mathcal{P}_2$.  We will $2$-privately compute a degree $2$ polynomial function $\phi({\bf X})$ with rate $(N-T)/N = 1/3$.  To identify polynomials with the functions they define as in Proposition \ref{polyfuns}, we assume that $|\K|\geq4$.  
	
	Suppose we want to evaluate some $\phi\in \mathcal{P}_2$ on the data vector ${\bf X}$.  Choosing a basis of $\mathcal{P}_2$ and sampling the $\psi_n\in \mathcal{P}_2$ as in (\ref{psin}), the queries are of the form
	\begin{equation}
	\rho_1 = \psi_1 + \phi, \ \
	\rho_2 = \psi_2,\ \
	\rho_3 = \psi_3
	\end{equation}
	any two of which are uniform on $\mathcal{P}_2^2$ and independent of the random variable $\mathsf{Q}$.  The response vector is of the form
	\begin{equation}
	\boldsymbol{\rho}({\bf X}) = \boldsymbol{\psi}({\bf X})+ \left[\phi({\bf X})\ 0 \ 0 \right] \in \K^{1\times 3}
	\end{equation}
	Right-multiplication by ${\bf H}_\mathcal{D}$ (the all-ones column vector of length $3$) yields $\boldsymbol{\rho}({\bf X})\cdot {\bf H}_\mathcal{D} = \phi({\bf X})$.
	\end{example}

	\subsection{Remarks on Upload Cost}
	
	Using the $T$-PC scheme of Section \ref{scheme}, every iteration the user must upload $N$ elements of $\mathcal{Q}$, which amounts of $NQ$ elements of $\F$.  On the other hand, the user receives $N$ elements of $\K$ in return, or equivalently $N\cdot [\K:\F]$ elements of $\F$.  Thus ignoring upload costs amounts to assuming that $[\K:\F]\gg \dim_\F\mathcal{Q}$.  So ignoring upload costs when evaluating high-degree polynomials thus requires the data matrix ${\bf X}$ to be defined over a very large extension of $\F$.

	\section{$T$-Private Computation of Systematically Encoded Data}
	
	In this section we consider polynomial computation of data encoded using a systematic storage code $\mathcal{C}$.  Our construction is again a generalization of that of \cite{FGHK16}, in that the user guarantees $T$-privacy by using an $[N,T]$ MDS retrieval code $\mathcal{D}$.  In the PIR scheme of \cite{FGHK16} the user recovers the desired file by decoding in $\mathcal{C}\star\mathcal{D}$.  When the query space is generalized to be $\mathcal{Q} = \mathcal{P}_G$ with $G\geq 1$, the user now recovers the desired function evaluations by decoding in $\mathcal{C}^{\star G}\star D$.  Reed-Solomon codes are shown to be especially useful in this context, in which case the download rate decreases linearly in $G$.
	
	\subsection{Scheme Construction}\label{main_construction}
	
	
	Let $\mathcal{Q}$ be an $\F$-linear query space, and define
	\begin{equation}
	\mathcal{C}^\mathcal{Q} = \text{span}_\K\{
	\left[
	\psi({\bf y}_1),\ldots,\psi({\bf y}_N)
	\right] \in \K^{1\times N}\ |\ \psi\in \mathcal{Q}
	\}
	\end{equation}
	where we range over all possible ${\bf Y} = [{\bf y}_1,\ldots,{\bf y}_N] = {\bf X}\cdot {\bf G}_\mathcal{C}$ for  ${\bf X}\in \K^{M\times K}$.  For example, if $\mathcal{Q} = \mathcal{P}_1$ then $\mathcal{C}^\mathcal{Q} = \mathcal{C}_\K$.

	We pick an $[N,T]_\F$ MDS code $\mathcal{D}$ as before, and define the $\psi_n\in \mathcal{Q}$ as in (\ref{psin}).  The first iteration of the scheme proceeds as follows.  Let $F\leq K$ be a constant whose exact value will be determined shortly.  We define the queries by setting
	\begin{equation}\label{queries}
	\rho_n = \left\{\begin{array}{cl}
	\psi_n + \phi_1 & \text{if $n\leq F$} \\
	\psi_n & \text{if $n > F$}
	\end{array}\right.
	\end{equation}
	so that the total response vector is of the form
	\begin{equation}\label{total_response_hard}
	\boldsymbol{\rho}({\bf Y})
	=
	\underbrace{\left[
	\psi_1({\bf y}_1),
	\ldots,
	\psi_N({\bf y}_N)
	\right]}_{=:\boldsymbol{\psi}({\bf Y})}
	+
	\left[
	\phi_1({\bf x}_1),
	\ldots,
	\phi_1({\bf x}_F),
	{\bf 0}
	\right]
	\end{equation}
	where ${\bf y}_k = {\bf x}_k$ for $k\leq F\leq K$ because $\mathcal{C}$ is in systematic form.    One easily computes that
	\begin{equation}
	\boldsymbol{\psi}({\bf Y}) = \sum_{q = 1}^Q
	\left[
	\psi^q({\bf y}_1),
	\ldots,
	\psi^q({\bf y}_N) 
	\right] \star {\bf d}^q
	\in \mathcal{C}^\mathcal{Q}\star \mathcal{D}_\K
	\end{equation}
	Let $\mathcal{E} = \mathcal{C}^\mathcal{Q}\star \mathcal{D}_\K$, let $D_\mathcal{E}$ be the minimum distance of $\mathcal{E}$, and choose $F=\min\{D_\mathcal{E}-1,K\}$.   Since every $D_\mathcal{E}-1$ rows of ${\bf H}_\mathcal{E}$ are linearly independent, right-multiplication of $\boldsymbol{\rho}({\bf Y})$ by ${\bf H}_\mathcal{E}$ allows us to decode the function values $\phi_1({\bf x}_1),\ldots,\phi_1({\bf x}_F)$ from (\ref{total_response_hard}).  This is simple error decoding in $\mathcal{E}$ with errors in prescribed positions, which can always correct up to $D_\mathcal{E}-1$ such errors.  Here the ``errors'' are the values $\phi_1({\bf x}_k)$.
	
	We iterate the above by writing $K = L\cdot F + \bar{F}$ with $0\leq \bar{F}< F$.  Suppose first that $\bar{F} = 0$, so that $F$ divides $K$.  To compute the rest of the function evaluations $\phi_1({\bf x}_{F+1}),\ldots,\phi_1({\bf x}_K)$ we use $L$ total iterations of the scheme, using the next successive set of $F$ servers each iteration when adding $\phi_1$ to the functions $\psi_n$ as in (\ref{queries}).  After $L$ iterations, we have thus computed $\phi_1({\bf x}_k)$ for all $1\leq k\leq K$, and if $B > 1$ we simply repeat the above process with each $\phi_b$ in place of $\phi_1$.  In the case where $\bar{F} = 0$ the scheme is complete.
	
        	\begin{figure}
        \centering
        \begin{center}
        $\left[\phi_b({\bf x}_k)\right]=$
        \begin{tabular}{|c|c|c|c|c|c|c|c|}
        \hline
        (1) & (1) & (1) & (1) & (1) & (1) & (2) & (2) \\
        \hline
        (2) & (2) & (2) & (2) & (3) & (3) & (3) & (3) \\
        \hline
        (3) & (3) & (4) & (4) & (4) & (4) & (4) & (4) \\
        \hline
        \end{tabular}
        \end{center}
        \caption{Visualizing the iteration process for the PC scheme via the matrix $\left[\phi_b({\bf x}_k)\right]\in \K^{3\times 8}$, for a system with parameters $K = 8$, $B = 3$, and $F =6$.  An integer $(s)$ in an entry of the above signifies that that entry is downloaded during the $s^{th}$ iteration.  We require $S = KB/F = 4$ iterations in total.}\label{scheme_fig}
        \end{figure}
	
	When $\bar{F}>0$ we first iterate the scheme $L$ times to compute $\phi_1({\bf x}_1),\ldots,\phi_1({\bf x}_{LF})$ as above.  In the $(L+1)^{st}$ iteration, we set
	\begin{equation}\label{wraparound}
	\rho_n = \left\{
	\begin{array}{cl}
	\psi_n + \phi_1 & \text{if $LF + 1\leq n \leq K$} \\
	\psi_n + \phi_2 & \text{if $1\leq n \leq F-\bar{F}$} \\
	\psi_n & \text{otherwise}
	\end{array}
	\right.
	\end{equation}
	and decode in $\mathcal{E}$ to download the function values $\phi_1({\bf x}_{LF+1}),\ldots,\phi_1({\bf x}_K)$, as well as the values $\phi_2({\bf x}_1),\ldots,\phi_2({\bf x}_{F-\bar{F}})$, for a total of $F$ values downloaded during this iteration.  The scheme now proceeds in an obvious way, downloading values $\phi_2({\bf x}_k)$ in groups of $F$ at a time until we get to the end of the list of systematic servers, where we ``wrap around'' and download values $\phi_3({\bf x}_k)$ starting with server $1$, as in (\ref{wraparound}) and as illustrated in Fig.\ \ref{scheme_fig}.  We continue this way until all values $\phi_b({\bf x}_k)$ are downloaded, choosing the block length $B$ so that $KB$ is divisible by $F$.  In this case, the scheme requires $S = KB/F$ iterations.  

	\begin{theorem}\label{maintheorem}
	With the notation as in the above scheme construction, assume that $F$ divides $KB$.  Then the above PC scheme is $T$-private and has rate $R = F/N$.
	\end{theorem}
	\begin{IEEEproof}
	The proof of $T$-privacy is exactly as in Theorem \ref{repthm} and is thus omitted.  The download rate is readily calculated to be $R = \frac{KB}{SN} = F/N$ as claimed.
	\end{IEEEproof}

	\subsection{$T$-Private Computation of Polynomial Functions of Systematically Encoded Data}
	
	In general, determining the code $\mathcal{C}^\mathcal{Q}$ and the minimum distance and parity-check matrix of $\mathcal{E}$ may be intractable for a general query space $\mathcal{Q}$.  However, when $\mathcal{Q} = \mathcal{P}_G$, the codes $\mathcal{C}^\mathcal{Q}$ and $\mathcal{E}$ take a particularly simple form.
	
	\begin{proposition}\label{polystar}
	Let $\mathcal{C}$ be a linear code such that $\mathcal{C}\subseteq\mathcal{C}\star\mathcal{C}$.  Then for the query space $\mathcal{Q} = \mathcal{P}_G$, we have $\mathcal{C}^\mathcal{Q} = (\mathcal{C}_\K)^{\star G}$ and therefore $\mathcal{E} = (\mathcal{C}^{\star G}\star \mathcal{D})_\K$.
	\end{proposition}
	\begin{IEEEproof}
	Picking a basis of $\mathcal{P}_G$ to consist only of monomials with $\F$-coefficients, we see that it suffices to show that $\boldsymbol{\psi}^q({\bf Y}) = [\psi^q({\bf y}_1),\ldots,\psi^q({\bf y}_N)]$ is in $(\mathcal{C}_\K)^{\star G}$ for any monomial $\psi^q$ of degree $\leq G$.  Suppose that $\psi^q = X_1^{a_1}\cdots X_M^{a_M}$, and write ${\bf y}_n = [y_{n1},\ldots,y_{nM}]^\mathsf{T}$.  Note that for each $m$, the vector $[y_{1m},\ldots,y_{Nm}]$ is a codeword in $\mathcal{C}_\K$.  We have
	\begin{align}
	\boldsymbol{\psi}^q({\bf Y}) &= 
	\left[
	\prod_{m = 1}^M y_{1m}^{a_m},\ldots,\prod_{m=1}^M y_{Nm}^{a_m}
	\right] \\
	&= \left[
	y_{11},\ldots,y_{N1}
	\right]^{\star a_1}
	\star
	\cdots
	\star
	\left[
	y_{1M},\ldots,y_{NM}
	\right]^{\star a_M} \nonumber \\
	&\in (\mathcal{C}_\K)^{\star a_1}\star\cdots\star (\mathcal{C}_\K)^{\star a_M} \subseteq (\mathcal{C}_\K)^{\star G} \nonumber
	\end{align}
	and hence $\boldsymbol{\psi}^q({\bf Y})\in (\mathcal{C}_\K)^{\star G}$ as claimed.  We have shown that $\mathcal{C}^\mathcal{Q}\subseteq(\mathcal{C}_\K)^{\star G}$, and equality follows by picking the $\psi^q$ to be all monomials of degree exactly $G$.
	\end{IEEEproof}

	

	Due to the simplicity of the star-product of RS codes as noted in (\ref{rsstar1}) and (\ref{rsstar2}), applying the above when $\mathcal{C}$ and $\mathcal{D}$ are Reed-Solomon codes yields explicit rate expressions.
	
	\begin{theorem}\label{grsrate}
	Let $\mathcal{C} = \mathcal{RS}_K(\boldsymbol{\alpha})$ and $\mathcal{D} = \mathcal{RS}_T(\boldsymbol{\alpha})$, for some evaluation vector $\boldsymbol{\alpha}\in \F^{1\times N}$.  Suppose that $G(K-1)+T\leq N$.  Then the download rate of the $T$-PC scheme for the query space $\mathcal{Q} = \mathcal{P}_G$ of the previous subsection is given by
	\begin{equation}
	R = \min\{N-(G(K-1)+T),K\}/N
	\end{equation}
	\end{theorem}
	\begin{IEEEproof}
	Note that all RS codes contain the repetition code, so the condition $\mathcal{C}\subseteq\mathcal{C}\star\mathcal{C}$ is satisfied.  One now calculates using Proposition \ref{polystar}, (\ref{rsstar1}), and (\ref{rsstar2}) that $\mathcal{E} = (\mathcal{C}_\K)^{\star G}\star \mathcal{D}_\K$ is an RS code with minimum distance $N-(G(K-1)+T)+1$. 
	\end{IEEEproof}
	
	\section{Conclusions and Future Work}
	
	In this paper we have generalized the methods of \cite{FGHK16} to construct $T$-Private Computation schemes.  The first scheme, for a data vector replicated over $N$ servers of which any $T$ can collude, achieves a rate of $(N-T)/N$ for arbitrary functions of the data.  This is same as the PIR rate achieved in \cite{FGHK16}.  The second scheme evaluates arbitrary functions of data encoded systematically with an $[N,K]$ storage code.  When the storage and retrieval codes are Reed-Solomon codes, we obtain a rate of $\min\{N-(G(K-1)+T),K\}/N$ for evaluating degree $G$ polynomial functions.
	
	We remark that when $G = 1$, an obvious variant of the scheme of Section \ref{main_construction} (which we save for an extended version of this paper) improves the rate to $(D_\mathcal{E}-1)/N$.  However, this improvement seems difficult for $G > 1$, as the function evaluations $\phi_b({\bf x}_k)$ cannot be readily obtained from values $\phi_b({\bf y}_n)$ when $\phi_b$ is non-linear and $n$ is a non-systematic node.  Making this improvement for higher-degree polynomials and general non-linear functions is our next immediate order of business.  Secondly, our scheme is easily symmetrizable by having the servers jointly add a uniform random codeword in $\mathcal{E}$ to the total response vector, a nice feature which we will incorporate into future work.  Lastly, many capacity achieving schemes \cite{sun_jafar_1,sunjafarpc} employ query spaces $\mathcal{Q}$ which are not $\F$-vector spaces.  Studying potential rate improvements given by non-linear query spaces is thus a worthwhile open problem.

	\bibliographystyle{IEEEtran}
	\bibliography{references}

\begin{thebibliography}{10}
\providecommand{\url}[1]{#1}
\csname url@samestyle\endcsname
\providecommand{\newblock}{\relax}
\providecommand{\bibinfo}[2]{#2}
\providecommand{\BIBentrySTDinterwordspacing}{\spaceskip=0pt\relax}
\providecommand{\BIBentryALTinterwordstretchfactor}{4}
\providecommand{\BIBentryALTinterwordspacing}{\spaceskip=\fontdimen2\font plus
\BIBentryALTinterwordstretchfactor\fontdimen3\font minus
  \fontdimen4\font\relax}
\providecommand{\BIBforeignlanguage}[2]{{%
\expandafter\ifx\csname l@#1\endcsname\relax
\typeout{** WARNING: IEEEtran.bst: No hyphenation pattern has been}%
\typeout{** loaded for the language `#1'. Using the pattern for}%
\typeout{** the default language instead.}%
\else
\language=\csname l@#1\endcsname
\fi
#2}}
\providecommand{\BIBdecl}{\relax}
\BIBdecl

\bibitem{PIR_original}
B.~Chor, E.~Kushlevitz, O.~Goldreich, and M.~Sudan, ``Private information
  retrieval,'' \emph{Journal of the ACM}, vol.~45, no.~6, pp. 965--981, 1998.

\bibitem{sun_jafar_1}
H.~Sun and S.~A. Jafar, ``The capacity of private information retrieval,''
  \emph{IEEE Transactions on Information Theory}, vol.~63, no.~7, pp.
  4075--4088, July 2017.

\bibitem{sun_jafar_2}
------, ``The capacity of robust private information retrieval with colluding
  databases,'' \emph{IEEE Transactions on Information Theory}, 2017.

\bibitem{bananaman}
K.~Banawan and S.~Ulukus, ``The capacity of private information retrieval from
  coded databases,'' 2016, arXiv: 1609.08138.

\bibitem{FGHK16}
R.~Freij-Hollanti, O.~W. Gnilke, C.~Hollanti, and D.~A. Karpuk, ``Private
  information retrieval from coded databases with colluding servers,''
  \emph{SIAM Journal on Applied Algebra and Geometry}, vol.~1, no.~1, pp.
  647--664, 2017.

\bibitem{patternISIT2017}
R.~Tajeddine, O.~W. Gnilke, D.~Karpuk, R.~Freij-Hollanti, C.~Hollanti, and
  S.~E. Rouayheb, ``Private information retrieval schemes for coded data with
  arbitrary collusion patterns,'' in \emph{2017 IEEE International Symposium on
  Information Theory (ISIT)}, June 2017, pp. 1908--1912.

\bibitem{zhang_ge_variants}
\BIBentryALTinterwordspacing
Y.~Zhang and G.~Ge, ``Private information retrieval from mds coded databases
  with colluding servers under several variant models,'' 2017. [Online].
  Available: \url{https://arxiv.org/abs/1705.03186}
\BIBentrySTDinterwordspacing

\bibitem{kumar_PIR_journal}
\BIBentryALTinterwordspacing
S.~Kumar, H.-Y. Lin, E.~Rosnes, and A.~G. i~Amat, ``Achieving private
  information retrieval capacity in distributed storage using an arbitrary
  linear code,'' 2017. [Online]. Available:
  \url{https://arxiv.org/abs/1712.03898}
\BIBentrySTDinterwordspacing

\bibitem{us_rm_pir}
\BIBentryALTinterwordspacing
R.~Freij-Hollanti, O.~Gnilke, C.~Hollanti, A.-L. Horlemann-Trautmann,
  D.~Karpuk, and I.~Kubjas, ``$t$-private information retrieval schemes using
  transitive codes,'' 2017. [Online]. Available:
  \url{https://arxiv.org/abs/1712.02850}
\BIBentrySTDinterwordspacing

\bibitem{bananamanbyz}
\BIBentryALTinterwordspacing
K.~Banawan and S.~Ulukus, ``The capacity of private information retrieval from
  byzantine and colluding databases,'' 2017. [Online]. Available:
  \url{https://arxiv.org/abs/1706.01442}
\BIBentrySTDinterwordspacing

\bibitem{sunjafarpc}
\BIBentryALTinterwordspacing
H.~Sun and S.~Jafar, ``The capacity of private computation,'' 2017. [Online].
  Available: \url{https://arxiv.org/abs/1710.11098}
\BIBentrySTDinterwordspacing

\bibitem{ma77}
F.~J. MacWilliams and N.~J.~A. Sloane, \emph{The Theory of Error-Correcting
  Codes}.\hskip 1em plus 0.5em minus 0.4em\relax Amsterdam: North Holland,
  1977.

\bibitem{miropc}
\BIBentryALTinterwordspacing
M.~Mirmohseni and M.~A. Maddah-Ali, ``Private function retrieval,'' 2017.
  [Online]. Available: \url{https://arxiv.org/abs/1711.04677}
\BIBentrySTDinterwordspacing

\end{thebibliography}
	\end{document}